\documentstyle[aps,preprint]{revtex}
\begin{document}
\title{Analysis and modeling of scale-invariance in plankton abundance}
\author{Jon D. Pelletier}
\address{Department of Geological Sciences, Snee Hall, Cornell University,
Ithaca, NY 14853 \\ phone: 607-255-8481 FAX: 607-254-4780 \\ email: 
pelletie@geology.cornell.edu}
\maketitle
\newpage
\begin{abstract}
The power spectrum, $S$, of horizontal transects of plankton abundance are often
observed to have a power-law dependence on wavenumber, $k$, with exponent close
to $-2$: $S(k)\propto k^{-2}$ over a wide range of scales. I present power
spectral analyses of aircraft lidar measurements of phytoplankton abundance 
from scales of 1 to 100 km. A power spectrum $S(k)\propto k^{-2}$ is obtained.
As a model for this observation, I consider a stochastic growth equation where
the rate of change of plankton abundance is determined by turbulent mixing,
modeled as a diffusion process in two dimensions, and exponential growth
with a stochastically variable net growth rate representing a fluctuating environment.
The model predicts a lognormal distribution of abundance and a power spectrum of
horizontal transects $S(k)\propto k^{-1.8}$, close to the observed spectrum.  
The model equation predicts that the power spectrum of
variations in abundance in time at a point in space is $S(f)\propto f^{-1.5}$
(where $f$ is the frequency). 
Time series analysis of local variations of phytoplankton and
zooplankton yield a power-law power spectrum with exponents $-1.3$ and 
$-1.2$, respectively from time scales of one hour to one year. 
These values are roughly consistent with the model prediction of $-1.5$. 
The distribution of abundances is nearly lognormal as predicted. 
The model may be more generally applicable than for the spatial distribution
of plankton.
I relate the model
predictions to observations of spatial patchiness in vegetation.  
\end{abstract}
\newpage
Levin (1992) has argued that the central problem in ecology is the 
determination of the factors controlling the spatial pattern of ecosystems at the
wide range of scales at which spatial heterogeneity is observed. Particularly 
intriguing is how movement and interactions at small scales, such as competition, dispersal, 
and local environmental variations, can
result in spatial pattern at larger scales.

Several generic mechanisms have been proposed to generate spatial variations in
abundance.
For instance, diffusive instabilities 
in multiple species models can generate complex spatial patterns (Okubo 1974,
Levin and Segel 1976). Roughgarden (1977) has argued that random fluctuations 
in resources necessary for reproduction may also generate complex, patchy spatial 
distributions. In this
paper we propose a model for spatial heterogeneity based on the second mechanism: dispersal in
a stochastic environment. The model is an extension of the 
classic equation studied by Kierstead and Slobodkin (1953) and Skellam (1951) with
diffusive dispersal and exponential growth with 
the net growth rate, or production efficiency, a stochastic variable in space and time. I report on the
model behavior in space and time and compare the predictions to observations  
of spatial and temporal variations in plankton abundance. 

Plankton systems are uniquely valuable for the analysis and modeling
of spatial variations in abundance because remote sensing techniques make
possible high-resolution measurements in space and time over a wide range
of scales.   
Power spectra of one-dimensional
transects of phytoplankton 
and zooplankton abundance have been generally
found to be a power-law function of wavenumber with an exponent close
to $-2$: $S(k)\propto k^{-2}$ where $S$ denotes the power spectrum
(Horwood 1978; Platt and Denman 1975; Mackas and Boyd 1979; Denman and Abbott 1988). There are
differences in the reported spectra. Platt and Denman (1975) argued that
a crossover exists in their data from a high wavenumber region with 
exponent $-2$ to a lower wavenumber region with exponent $-\frac{5}{3}$.
The transition occurs at a a spatial scale of about 10 m. Horwood (1978)
found no such crossover. He also analyzed 
variations on much larger scales, up to 20 km. He found that the 
power spectral behavior with exponent $-2$ continued up to these scales.  
This is consistent with the results of Denman and Abbott (1988) who found 
$S(k)\propto k^{-2}$ from scales of 1 to 100 km with ocean color data.

I have carried out power spectral analyses of phytoplankton abundance 
estimated with measurements taken by the Airborne Lidar Observatory. 
I used 23 transects, about half of which were sampled in the continental shelf
off the coast of Maryland and half in the Arabian Sea. The power spectrum of 
each transect was estimated as the modulus squared of the Fast Fourier Transform.
Each of the spectra were then averaged at equal frequency values. The resulting
average spectrum is presented in Figure 1. The straight line in the plot represents
the form of the power-law spectrum $S(k)\propto k^{-2}$. A good fit is obtained 
except for the largest scales where the variance falls below the scale-invariant
trend. I found no significant difference in spectra between transects parallel
and perpendicular to the coast. These results are consistent with the results
of power spectral analysis of Denman and Abbott (1988) who found a $S(k)\propto k^{-2}$ 
spectrum over the same scales with ocean color data.    

Several models have been proposed to explain this spectrum.
Denman, Okubo, and Platt (1977) proposed
three different regimes for the power spectrum based on the relative 
importance of turbulent mixing and the rate of reproduction. Their spectrum
predicts that the power spectrum should decrease sharply below a characteristic
wavenumber. The continual increase of the power spectrum at decreasing
wavenumbers observed in my analysis and the others I have cited
appears to invalidate this prediction. 

Fasham (1978) has
simulated a number of stochastic models that give rise to power spectra 
similar to $S(k)\propto k^{-2}$. Fasham begins by presenting
an extension of the equation first studied by Kierstead and 
Slobodkin (1953) and Skellam (1951) representing the abundance of a species
in time and space with turbulent dispersion modeled by the diffusion equation 
and exponential growth with a constant growth rate. 
The original equation in one spatial dimension is  
\begin{equation}
\frac{\partial a}{\partial t}=D\frac{\partial^{2}a}{\partial x^{2}} +\alpha a
\end{equation}
where $a$ is the local abundance and $\alpha$ is the growth rate.
Fasham (1978) extended this equation by including a Gaussian white noise
$\eta (x,t)$ to represent a fluctuating environment. The resulting
equation is
\begin{equation} 
\frac{\partial a}{\partial t}=D\frac{\partial^{2}a}{\partial x^{2}}
+\alpha a + \eta (x,t)
\end{equation} 
This equation generates a spatial distribution with a power spectrum
$S(k)\propto (Dk^{2}+|\alpha|)^{-1}$. At high wavenumbers this power
spectrum is $S(k)\propto k^{-2}$, similar to the spectrum often observed. 
However, this equation describes dynamics in one 
dimension only. As such, it is an inappropriate model for dynamics
of phytoplankton on the ocean surface where a two or three dimensional model
is appropriate. The above equation, generalized to two dimensions, has been 
used to model the height of a surface on which grains of sand are randomly 
deposited in space and time. The power spectrum of one-dimensional transects
of the surface generated by the model is not a power-law function (Edwards and Wilkinson 1982).
Fasham's model therefore does not explain the $S(k)\propto k^{-2}$ spectrum.
 
In this paper I will consider another extension of the diffusion
equation with exponential growth. In the model I include the effects of
a fluctuating environment by making the net growth rate or production
efficiency 
a random function 
in space and time. Allowing the production efficiency to vary may better reflect
the effects of a variable environment than a random forcing term uncoupled
to the abundance. 
Many complex processes are thought to affect the production efficiency
of phytoplankton: nitrate upwelling and downwelling (Powell and Richerson 1985),
fluctuations in light 
intensity caused by internal waves (Abbott, Powell, and Richerson 1982), and 
temperature fluctuations (Denman 1976). 
In this paper I will model environmental
variations resulting in variations in production efficiency
in the simplest manner possible: as a Gaussian white noise in space and time.
I consider the equation in two spatial dimensions,
as applicable for plankton 
on the ocean surface.
The model equation is
\begin{equation} 
\frac{\partial a}{\partial t}=D\nabla ^{2}a +\eta(x,y,t)a
\end{equation}
I will compare abundances generated by this equation to previous studies
of spatial heterogeneity in plankton and my 
analysis of phytoplankton heterogeneity in space and time with remotely sensed
abundance.

The above equation has been well studied in the physics literature where it
is a variant of the Kardar-Parisi-Zhang (KPZ) equation (Kardar, Parisi, and Zhang 1986).
The KPZ equation was originally proposed to model the fractal
characteristics of atomic surfaces grown by ion deposition. The KPZ
equation in two spatial dimensions for a surface with elevation $h(x,y,t)$ is
\begin{equation}
\frac{\partial h}{\partial t}=D\nabla ^{2}h+\lambda (\nabla h)^{2}+
\eta (x,y,t)
\end{equation}
The surface generated by this equation has one-dimensional transects with 
a Gaussian distribution and power
spectra $S(k)\propto k^{-1.8}$ (Amar and Family 1990, Moser, Wolf, and Kertesz 1991).
The model also has local variations in time with a power spectrum $S(f)\propto f^{-1.5}$.
With the transformation $a=exp(\frac{\lambda}{D}h(x,y,t))$ the above equation becomes
(Kardar, Parisi, and Zhang 1986)
\begin{equation} 
\frac{\partial a}{\partial t}=D\nabla ^{2}a +\frac{\lambda}{D}\eta(x,y,t)a
\end{equation}
which is the same as equation (3) except for the multiplicative factor in the
stochastic term. This factor does not change the form of the power spectra since
the spectra are independent of the variance of the noise term $\eta(x,y,t)$.

The spectral behavior of the solution to the Kardar-Parisi-Zhang equation can be
used to determine the behavior of equation (5). Since the Kardar-Parisi-Zhang equation
generates a Gaussian distribution and equation (5) has been obtained by an
exponential transformation, the distribution of abundances generated by equation (5)
will be lognormal. The power spectra of variations in abundance in space and time 
will be the same as those of the Kardar-Parisi-Zhang equation. This conclusion is
based on the power spectral analyses of Gomes da Silva and Turcotte (1994) with 
Gaussian and lognormal noises with power-law power spectra. These authors found
that the form of the spectra are unchanged after an exponential transformation unless
the data are extremely skewed. The power spectrum of horizontal transects, 
$S(k)\propto k^{-1.8}$ is close to the observed spectrum with an exponent close to
$-2$ from scales of meters to hundreds of kilometers.

To test the model prediction of local variations in time, we have obtained time series
of local phytoplankton and zooplankton abundance off the coast of Maryland sampled 
hourly up to time scales of one year from Wirick (1996). The data series are described
in Ascioti et al. (1993) where the authors tested the hypothesis that chaos may explain
the aperiodic fluctuations in local abundance observed in the time series. Although
their results did not conclude the presence of chaos, they suggested that nonlinear
dynamics may account for at least some of the variability in the data. In Figure 2
we present the cumulative distribution function (thick lines) of the data to test the
model prediction of a lognormal distribution. The thin lines represent the cumulative
lognormal distribution fit to the data. Reasonably good fits are obtained. The
power spectra of each of the series are presented in Figure 3. Both series exhibit
a good fit to a power-law relationship with exponents -1.3 and -1.2 for phytoplankton
and zooplankton, respectively. These values are close to the exponent -1.5 predicted by
the model equation (5).

The model I have introduced may be more generally applicable than for plankton.
The terms in the model equation (5) may describe aspects of growth and dispersal 
applicable to terrestrial species. Diffusion is a common parameterization
for dispersal in terrestrial species. Karieva (1983) has used the diffusion equation
to model insect dispersal. Hengeveld and Haeck (1982) have analyzed the spatial 
distribution of variety of species and found that species generally have a maximum
abundance at the center of their range with abundance tapering off at the edges of
the range. A diffusion model of dispersal or migration is consistent with
this general biogeographical rule and is one of the models the authors propose to
explain their observations (Hengeveld and Haeck 1981). In the first application
of the diffusion model to dispersal, Skellam (1951) applied the model to the colonization
of species as diverse as muskrats and oak trees. The other term in the equation,
exponential growth in a fluctuating environment, may also be generally applicable for
``opportunistic'' species for which the environmental fluctuations are strong enough
that the species abundance rarely approaches the local carrying capacity. For species
whose abundance is often close to the local carrying capacity, exponential growth
is inappropriate and the logistic growth terms must be used. 
The model we have studied predicts a lognormal distribution of abundance as generally 
observed for species in a wide range of habitats (Preston 1962, May 1975). A number
of quantitative studies of patchiness in vegetation have been performed. The results
of these investigations
are consistent with the spatial power spectrum $S(k)\propto k^{-2}$. These
will be discussed after a brief discussion of fractal geometry.

Topography is a useful analog for the variations in the
abundance of phytoplankton and 
zooplankton on the ocean surface. The analogy is useful for introducing fractal aspects
of the model and their possible relation to vegetation patchiness analyzed with fractal
measures. The power spectrum of topography
generally has a power-law dependence $S(k)\propto k^{-2}$ 
(Fox and Hayes 1985) as observed for plankton abundance.   
Topography is also a type example of a self-affine or fractal function.
Introductory geology students learn that objects (a person, a hammer,
or a coin) must be placed into many geological photographs 
in order to set the scale because topographic variations
have no characteristic scale. An introduction to the application of fractal geometry
in ecology is given by Sugihara and May (1990).   
There are three fractal aspects of the Earth's topography. 
The Earth's surface 
is generally self-affine with Hausdorff measure $Ha=\frac{1}{2}$.
The Hausdorff measure is defined
by the relationship between 
the change in elevation, $\Delta h$, in the case of topography,
over a horizontal distance 
interval, $\Delta x$: $\Delta h\propto\Delta x^{Ha}$. 
Ahnert (1984) has found $Ha \approx \frac{1}{2}$ from topographic maps. A random walk is another 
example of a function with Hausdorff dimension $Ha=\frac{1}{2}$. Another measure of the fractal
geometry of topography is given by the exponent of the power spectrum of one-dimensional
transects. This measure, $\beta$ is usually defined by the negative of the power spectral
exponent: $S(k)\propto k^{-\beta}$. For topography and plankton abundance, 
$\beta\approx 2$.
$Ha$ and $\beta$ are related by the relation $\beta = 1+2Ha$
(Turcotte 1992). The second fractal aspect of topography is that 
coastlines and contour lines are often fractal with dimension 1.25.
On water-filled topography, a coastline is the connected subset of the
topography at the constant elevation of sea level. Finally, the cumulative number of islands, 
domains of topography above a fixed elevation, greater than an
area, $A$, has a 
power-law dependence on area with an exponent close 
to $-\frac{3}{4}: N(>A)\propto A^{-\frac{3}{4}}$ (Korcak 1938).
A similar area distribution is observed for lakes (Maybeck 1995). 
Each of these fractal characteristics of topography have been identified
in species abundance patterns. The power spectral analysis
of phytoplankton and zooplankton has been discussed. Krummel et al. (1987)
has calculated the fractal dimension of deciduous forest patches in the 
U. S. Geological Survey Natchez, Mississippi 1:250,000 Quadrangle. Although 
they emphasized the variation in perimeter fractal dimension between
different scales, the fractal dimension varies only between about 1.2
for small scales and 1.4 for the larger scales. Given the statistical
variation in the fractal dimension, these values are not inconsistent with  
a value of $1.25$ as measured for coastlines. The cumulative frequency-area
distribution of Cypress and Broadleaf patches in the Okefenokee Swamp
was calculated by Hastings et al. (1982). They obtained exponents in
the range of $-0.5$ to$-0.8$, similar to the exponent obtained in the 
cumulative distribution of islands and lakes. Collins and Glenn (1990) have identified
similar scale-invariant or hierarchical
patchiness in grasslands. 
Kondev and Henley (1995) have shown that two-dimensional fractal functions 
have fractal
``coastlines'' (defined as subsets of constant magnitude) and a power-law 
cumulative distribution of areas above a threshold magnitude (``islands'').
They showed that there is a unique relationship between the fractal dimension
of ``coastlines,'' the size distribution of ``islands,'' and the exponent
$\beta$ in the power spectrum. As a result, similar values of the fractal
dimension of ``coastlines'' and the size distribution of ``islands'' for
topography and vegetation patches imply that the power spectra of variations
in topography and species abundance along a transect are comparable. 
This suggests that the power spectrum of
abundance variations in vegetation for the study areas above may be roughly
consistent with the model power spectrum $S(k)\propto k^{-2}$. 
Palmer (1988) has performed variogram analysis of
spatial variations in abundance in vegetative species. 
Although his results imply that the negative of the power spectral exponent, $\beta$,
is generally smaller than $2$ and exhibits considerable variability
between species and habitats,
he concluded that
fractal geometry is a useful description of spatial variations in vegetative
abundance. I hope that this work stimulates the continued study of species
abundance patterns with fractal geometry with particular emphasis on the
universality of the fractal measures obtained.   

\section{acknowledgements}
I wish to thank the Airborne Lidar Observatory for making their data 
available over the internet and 
Creighton Wirick of Brookhaven National
Laboratory for generously sharing
his time series of local variations in phytoplankton and zooplankton 
abundance with me.

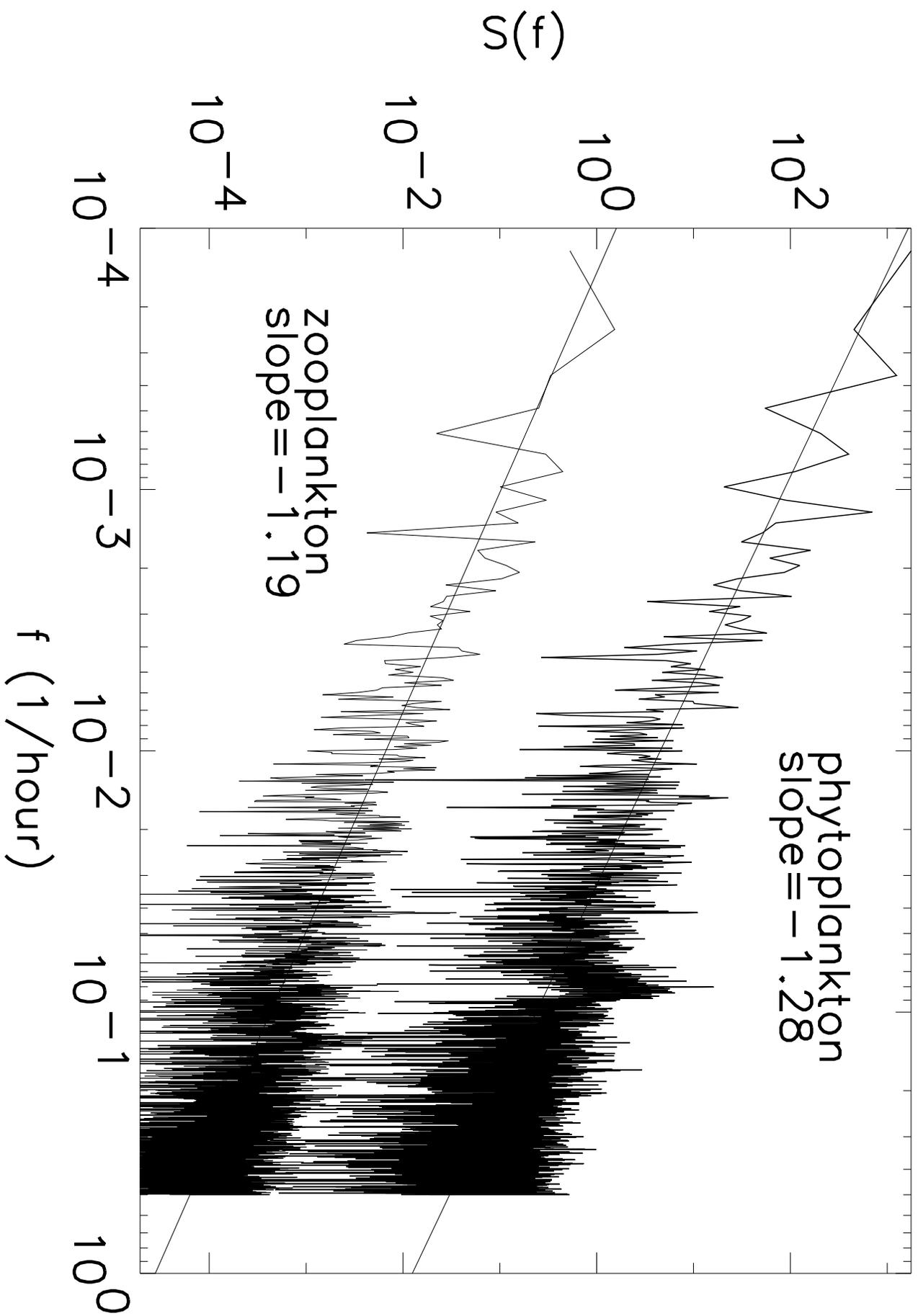
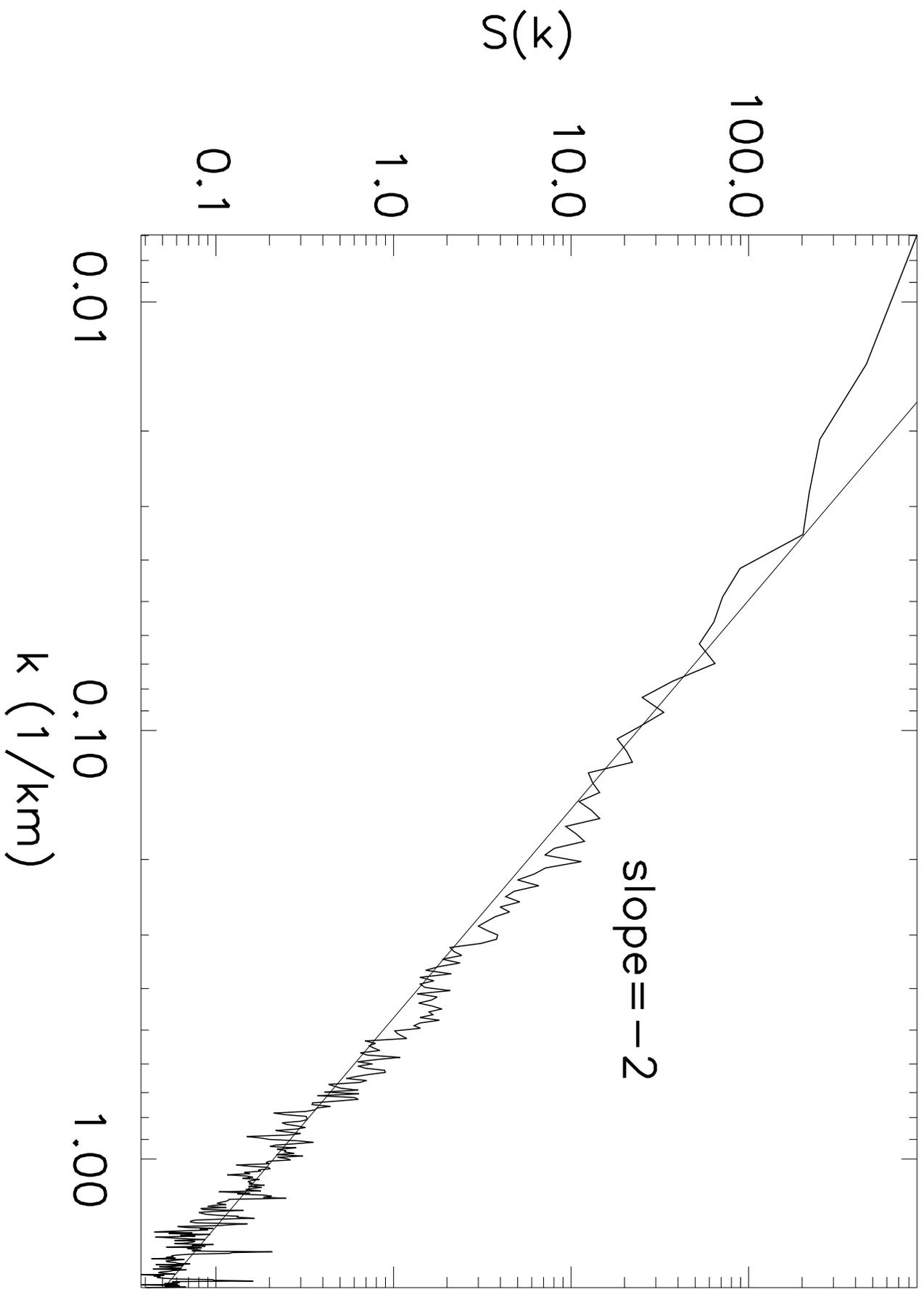
\end{document}